\documentclass[preprint, superscriptaddress,amsmath, nofootinbib]{revtex4}
\usepackage{graphicx}
\usepackage{latexsym}
\usepackage{slashed}
\usepackage{bm}
\usepackage{color}
\usepackage[colorlinks,citecolor=blue,urlcolor=blue,linkcolor=blue]{hyperref}
\usepackage{diagbox}

\def\({\left(}
\def\){\right)}

\def\beq{\begin{equation}}
\def\eeq{\end{equation}}

\begin{document}

\title{Probing Light DM through Migdal Effect with Spherical Proportional Counter}
	
\author{Yuchao Gu}
\email{guyc@njnu.edu.cn}
\affiliation{Department of Physics and Institute of Theoretical Physics, Nanjing Normal University, Nanjing, 210023, China}

\author{Jie Tang}
\email{tangjie@njnu.edu.cn}
\affiliation{Department of Physics and Institute of Theoretical Physics, Nanjing Normal University, Nanjing, 210023, China}
\affiliation{School of Physics, Southeast University, Nanjing 211189, China}

\author{Lei Wu}
\email{leiwu@njnu.edu.cn}
\affiliation{Department of Physics and Institute of Theoretical Physics, Nanjing Normal University, Nanjing, 210023, China}

\author{Bin Zhu}
\email{zhubin@mail.nankai.edu.cn}
\affiliation{Department of Physics, Yantai University, Yantai 264005, China}

\begin{abstract}
Owing to its low electronic noise and flexible target materials, the Spherical Proportional Counter (SPC) with a single electron detection threshold can be utilized to search for sub-GeV dark matter (DM). In this work, we investigate the prospects for light DM direct detection through the DM-nucleus Migdal effect in the DARKSPHERE detector. We consider the different DM velocity distributions and momentum-transfer effects. For Xenon and Neon targets, we find that the DM mass $m_{DM}$ can be probed down to as low as $m_{DM} \sim \mathcal{O}$(10) MeV, and the derived bounds on the DM-nucleus scattering cross section $\bar{\sigma}_{n}$ are sensitive to the high-velocity tails of the DM velocity distribution, which can be altered by orders of magnitude for the different DM velocity distributions in the region $m_{DM} < 10$ MeV. 
\end{abstract}

\maketitle

\tableofcontents

\newpage
\section{INTRODUCTION}
\label{sec1}
The existence of dark matter (DM) in the Universe has been confirmed by various cosmological and astrophysical observations. The Weakly Interacting Massive Particle (WIMP) with the typical mass at weak scale~\cite{Lee:1977ua, Jungman:1995df} has naturally correct thermal relic density and becomes the prevalent DM candidate. However, the current null results of searching for WIMPs in collider and (in)direct detection experiments have put strong bounds on many WIMP DM models~\cite{Bertone:2004pz, Feng:2010gw, Baudis:2014naa, Roszkowski:2017nbc, Bauer:2017qwy}. It is therefore essential to investigate other possibilities and develop new detection technologies to explore the light DM with mass from keV to GeV~(see recent reviews, e.g.~\cite{Essig:2011nj, Battaglieri:2017aum, Kahn:2021ttr}).

However, due to the nuclear recoil signals induced by light DM-nucleus scattering well below the threshold of DM direct detection experiments, light DM direct detection requires experiments with a lower threshold. Among these experiments, the New Experiments with Spheres-Gas (NEWS-G) using a Spherical Proportional Counter (SPC) can reach an extremely low energy threshold. It consists of a grounded copper shell and an anode in the center~\cite{Giomataris:2003bp, Bougamont:2010mj, NEWS-G:2017pxg, Katsioulas:2018squ, NEWS-G:2019lqz}, where a high-voltage electric field exists between the anode and the metal casing. The cavity can be filled with gas target materials, such as Helium, Neon, Argon, and Xenon. The main working principle of the SPC is that the DM arriving in the cavity may cause electron ionization via the interaction between DM and the target material. The ionized electrons drift toward the anode under the action of a high-voltage electric field and can eventually be detected by the sensor, which is connected to the anode. The first detector of the NEWS-G collaboration located in the Laboratoire Souterrain de Modane, with a diameter of 60 cm, has achieved the excellent exclusion limits on the DM-nucleus cross section $\bar{\sigma}_{n}$ for the DM mass being around 0.5 GeV~\cite{NEWS-G:2017pxg}. 

In this work, we investigate the prospects of detecting sub-GeV DM through the Migdal effect in the NEWS-G experiment with 3-meter-diameter DARKSPHERE detector, which is initially scheduled to operate in 2025. Different from the conventional DM direct detection experiments, the Migdal effect with lower threshold can play an important role in light DM direct detection~\cite{Ibe:2017yqa,Dolan:2017xbu,LUX:2018akb,EDELWEISS:2019vjv,Bell:2019egg,CDEX:2019hzn,Baxter:2019pnz,Essig:2019xkx,Su:2020zny,Knapen:2020aky,Liang:2020ryg,Liu:2020pat,Smith:2006ym,Flambaum:2020xxo,Wang:2021oha}. Since the first phase of the NEWS-G experiment packed with Neon can produce the competitive limits, we will compare the performance of Xenon and Neon as target materials.
Besides, the DM velocity distribution also has a significant impact on DM direct detection~\cite{Kuhlen:2009vh,McCabe:2010zh,Green:2017odb,Nunez-Castineyra:2019odi,Wang:2019jtk,Radick:2020qip,Wang:2021nbf,Gu:2022vgb}. Therefore we also study the dependence of the events generated by Migdal effect on three different DM velocity distributions and obtain the exclusion limits on $m_{DM}-\bar{\sigma}_{n}$ panel with regard to four different DM form factors $F_{DM}$.

The structure of our paper is organized as follows: In Sec.~\ref{sec2}, we recapitulate the computational framework of the Migdal effect in DM direct detection. In Sec.~\ref{sec3}, we discuss three different DM velocity distribution models and their possible impacts on the calculation results. In Sec.~\ref{sec4} we calculate differential event rates and give a comparison of differential event rates under various DM velocity distribution models. We will show our expected exclusion limits on the $m_{DM}-\bar{\sigma_{n}}$ plane. Finally, we will draw some conclusions in Sec.~\ref{sec5}.

\section{DM-NUCLEUS MIGDAL SCATTERING}
\label{sec2}
We start by introducing the Migdal effect differential cross-section for nuclear recoil energy $E_{R}$ and electron recoil energy $E_{e}$\cite{Ibe:2017yqa,bell2023exploring,Berghaus_2023,Li_2023,50collaboration2023search,nakamura2020detection,Liu_2020,Dey_2020,di_Cortona_2020},
\begin{equation}
\label{dsigmadER}
    \frac{d\sigma}{dE_R dE_{e}}\simeq \frac{1}{32\pi}\frac{m_N}{\mu_N^2 v_{DM}^2}\frac{|F_N(q)|^2|M(q)|^2}{(m_N +m_{DM})^2}\frac{d}{dE_{e}}\sum_F|Z_{FI}(q_e)|^2,
\end{equation}
with the invariant amplitude squared
\begin{equation}
    M(q)^2=(f_n (A-Z)+Z f_p)^2\times \overline{M(q)^2}.
\end{equation}
where $\mu_{N} = m_{DM}m_{N}/\left({m_{DM} + m_{N}}\right)$ is the DM-nucleus reduced mass and $v_{DM}$ is the velocity of the incoming DM particles. $|F_N(q)|^2$ is nuclear form factor. $q$ and $q_{e}$ are the nuclear and electron transfer momentum. A and Z are the mass number of atoms and the atomic number, respectively. $f_{p}$ ($f_{n}$) is dimensionless couplings of DM interacting with proton (neutron). $\overline{M(q)^2}$ represents the amplitude of DM scattering off a free nucleon. $Z_{FI}(q_e)$ is electron cloud transition factor, related to the electron ionization/excitation possibility. $\sum_F$ represents the sum of all possible final states wave functions of electrons. Note that the Migdal effect differential cross-section depends on the DM-nucleus elastic cross-section and the electron cloud transition factor. We rewrite the Migdal effect differential cross-section by exploiting the commonly defined reference cross section $\bar{\sigma}_{n}$ and DM form factor $|F_{DM}(q)|$,  
\begin{equation}
    \frac{d\sigma}{dE_R dE_{e}}\simeq \frac{1}{2}\frac{m_N}{\mu_N^2 v_{DM}^2} (f_n (A-Z)+Z f_p)^2 \bar{\sigma}_{n} |F_N(q)|^2 |F_{DM}(q)|^2 \frac{d}{dE_{e}}\sum_F|Z_{FI}(q_e)|^2
\end{equation}
with 
\begin{equation}
    \bar{\sigma}_n=\frac{\overline{|M(q=q_0)|^2} \mu_N^2}{16\pi m_{DM}^2 m_N^2},
\end{equation}
\begin{equation}
    |F_{DM}(q)|^2=\frac{\overline{|M(q)|^2}}{\overline{|M(q=q_0)|^2}},
\end{equation}
where $q_{0}=\alpha m_{e}$ is the reference momentum. The information related to transfer momentum $q$-dependent has been absorbed in DM form factor. Notably, the DM form factor $|F_{DM}(q)|$ equals to 1 ($\left(\alpha m_{e}/q\right)^2$) for a light (heavy) mediator.

In the following, we will pay more attention on the electron cloud transition factor $Z_{FI}(q_e)$\cite{Ibe:2017yqa,Bell_2022} described by
\begin{equation}
\label{Z2}
    \sum_F |Z_{FI}|^2=|Z_{II}|^2+\sum_{n,l,n',l'}p_{q_e}^d(nl\to n'l')+\sum_{n,l}\int\frac{dE_e}{2\pi}\frac{d}{dE_e}p_{q_e}^c(nl\to E_e),
\end{equation}
where $\{n,l\}$ and $\{n',l'\}$ are the energy levels of the electrons bound to target atom before and after scattering. $p_{q_e}^c$ and $p_{q_e}^d$ are the ionization and excitation probabilities of electrons. $|Z_{II}|^2$ represents the probability that the electrons are not affected by the DM-nucleus scattering process. The second term implies the probability of exciting the electrons from energy level $\{n,l\}$ to $\{n',l'\}$, while the third term is the probability of ionizing the electrons bound to energy level $\{n,l\}$ after DM-nucleus scattering. The ionization and excitation probabilities are discussed in detail in Refs~\cite{Ibe:2017yqa}, which reveals that the possibility of ionization can be orders of magnitude larger than that of excitation. In our following calculation, we will simply consider the ionization probability. 
By defining the term related to the ionization probability
\begin{equation}
    \frac{dp_{nl\to E_e}}{d{\rm ln} E_e}=\frac{\pi}{2}|f_{nl}^{ion}(E_e,q_e)|^2,
\end{equation} 
we can derive the differential cross section of Migdal effect
\begin{equation}
\frac{d\sigma}{dE_R dE_{e}}=\frac{\bar{\sigma}_n m_N (f_n(A-Z)+Zf_p)^2}{8\mu_N^2 v_{DM}^2}|F_N(q)|^2|F_{DM}(q)|^2\sum_{n,l} \frac{|f^{ion}_{n,l}(E_e,q_e)|^2}{E_e}.
\end{equation}
where $|f^{ion}_{n,l}(E_e,q_e)|$ is the ionization factor and $q_{e} \simeq \frac{m_{e}}{m_{N}} q$ is the momentum of electrons after scattering. By approximating the ionized electrons with Cowan's Hartree-plus-statistical-exchange method~\cite{Hamaide:2021hlp}, the ionization function $|f^{ion}_{n,l}(E_e,q_e)|$ can be written as

\begin{equation}
|f^{ion}_{n,l}(E_e,q)|^2= \left \langle\int\mathrm{d}\Omega_{k_e}\frac{2k^3_e}{8\pi^3}   
\times|\int\mathrm{d^3}x \psi_f^*(\boldsymbol{x},\boldsymbol{k_e})e^{i\boldsymbol{q}\cdot\boldsymbol{x}}\psi_i(\boldsymbol{x})|^2 \right\rangle,
\end{equation}
where $k_{e}=\sqrt{2m_e E_e}$ is the momentum of the unbound electron after scattering. The angled brackets indicate that the uniformly averaging over all orientations of the atom is taken into consideration.

In order to achieve the differential events as a function of $E_{e}$, we should integrate over the nuclear recoil energy $E_{R}$, which can be replaced by integrating over the transfer momentum $q$. For a given energy level $\{n,l\}$, the expression for the differential event rates as a function of $E_{e}$ with units $\rm{ton}^{-1}\cdot {\rm year}^{-1}\cdot {\rm keV}^{-1}$ can be written as~\cite{Essig:2019xkx,Essig:2011nj, Bringmann:2018cvk,Liao_2021,flambaum2023new}
\begin{equation}
\label{eq1}
    \frac{dR_{n,l}}{dE_e}=N_T\frac{\rho_{DM}}{m_{DM}}\frac{d\left\langle\sigma_{n,l}^{ion}v\right\rangle}{dE_e},
\end{equation}
with
\begin{eqnarray}
\label{eq2}
\frac{\mathrm{d}\left\langle \sigma _{n,l}^{ion} v\right\rangle}{\mathrm{d}E_e}&=&\frac{\bar{\sigma }_n}{8\mu _n^2 E_e}(f_n (A-Z)+Z f_p)^2\int_{q_{-}}^{q_{+}}\mathrm{d}q \nonumber \\ 
&\times& \Bigg[q|F_{DM}(q)|^2|F_N(q)|^2| f_{nl}^{ion}(E_e,q_e)|^2\eta (v_{min }(q,\Delta E_{n,l}))\Bigg] ,
\end{eqnarray}
and
\begin{equation}
    \eta(v_{min}(q,\Delta E_{n,l}))=\int_{v_{min}(q,\Delta E_{n,l})}^{v_{max}}\frac{f(v_{DM})}{v_{DM}}dv_{DM}.
\end{equation}
where $N_T$ is number density of target materials and $\rho_{DM} =0.3$ GeV/cm$^3$ is the energy density of the halo DM~\cite{Kapteyn:1922zz,Bahcall:1983uu,Dehnen:1996fa,McMillan:2011wd,Garbari:2012ff,Bovy:2012tw,Zhang:2012rsb,Read_2014}. $\Delta E_{n,l}$ is deposited energy of the electron and the $\eta (v_{min }(q,\Delta E_{n,l}))$ is the usual velocity average of the inverse speed. $f(v_{DM})$ is the velocity distribution of DM. Since the differential event rates depend on $f(v_{DM})$~\cite{Radick:2020qip}, we will take into account different velocity distribution models including Standard Halo Model(SHM)~\cite{Drukier:1986tm, OHare:2018trr, Evans:2018bqy, OHare:2019qxc, Buch:2020xyt}, Tsallis Model(Tsa)~\cite{Tsallis:1987eu,Hansen_2006,Vergados:2007nc,Ling:2009eh} and Empirical Model(Emp)~\cite{Mao:2012hf,Mao:2013nda,Bozorgnia:2017brl,Vogelsberger:2019ynw} in next section~\ref{sec3}.
Besides, the incoming DM velocity $v_{DM}$ satisfies the energy-momentum conservation conditions,
\begin{equation}
    \Delta E_{n,l}= \frac{1}{2} m_{DM} v_{DM}^2 -\frac{|m_{DM} \bf{v}-\bf{q}|^2}{2 m_{DM}}-\frac{\bf{q}^2}{2 m_{N}}
\end{equation}
where $\Delta E_{n,l}=E_e +|E_{n,l}|$ and $|E_{n,l}|$ is the bound energy of $\{n,l\}$ state. For a given transfer momentum $q$ and electron recoil energy $E_{e}$, we can derive the minimum velocity of the incoming DM particle\cite{qiao2023diurnal},
\begin{equation}
\label{vmin}
    v_{min}(q,\Delta E_{n,l})=\frac{q}{2 \mu_{N}}+\frac{E_e +|E_{n,l}|}{q}.
\end{equation}
Additionally, $q_{-}$ and $q_{+}$ can be obtained by Eq.~\ref{vmin},
\begin{equation}
\label{q}
    q_{\mp}=\mu_{N} v_{max} \left(1 \mp \sqrt{1-\frac{E_e + |E_{n,l}|}{\frac{1}{2} \mu_{N} v_{max}^2}}\right).
\end{equation}

By performing Taylor expansion on Eq.~\ref{q}, we can derive the minimum of the transfer momentum $q_{-}$
\begin{equation}
    q_{-}\sim \frac{E_e +|E_{n,l}|}{v_{max}}.
\end{equation}
Because of the binding energy of Xenon 5p $|E_{5p}|\sim$ 12.7 eV and $v_{max}\sim$ 760 km/s, the transfer momentum $q_{-}$ is typically at keV scale. While $q_{e}$ is well below 1 keV due to $q_e \simeq m_{e} q/m_{N}$ being highly suppressed by $m_{e}/m_{N}$. Therefore, we assume the dipole approximation for the ionization function in $q_{e}<1$ keV region, which is expressed by
\begin{equation}
\label{approximation}
    |f^{ion}_{n,l}(E_e,q_e)|^2=\frac{q_e^2}{(1~{\rm keV})^2}|f_{ion}^{i\to f}(E_e,q_e=1 ~{\rm keV})|^2   (q_e<1 ~{\rm keV}).
\end{equation}

\section{DM VELOCITY DISTRIBUTION}
\label{sec3}
The velocity distribution of DM plays a crucial role in the direct detection of DM. Previous studies have pointed out that the different velocity distributions will impact greatly the DM-electron scattering process. As discussed in reference[36], DM-electron scattering relies on the various velocity distributions of the DM halo, particularly on their significantly different high-velocity tails. Because the Migdal effect is related to the electron recoil energy, we consider the various velocity distributions in the Migdal effect, which may improve the limits on the DM-nucleus cross-section from the NEWS-G experiment. Besides, we first consider the impact of the Migdal effect on the sub-GeV DM direct detection by the NEWS-G experiment with DARKSPHERE detector. Therefore, we have investigated the dependence of the different DM velocity distribution models on the Migdal effect. Firstly, the most widely used model for the DM direct detection is the Standard Halo Model where the DM velocity distribution  follows a Maxwell-Boltzmann distribution in the Earth frame and is described by
\begin{equation}
    f_{SHM}(\vec v)=\frac{1}{K}e^{-|\vec v +\vec v_E|^2/v_0^2}\Theta(v_{esc}-|\vec v +\vec v_E|),
\end{equation}
with the normalization coefficient
\begin{equation}
    K=v_0^3 \left(\pi^{\frac{3}{2}} {\rm erf}(\frac{v_{esc}}{v_0})-2\pi \frac{v_{esc}}{v_0}e^{-\frac{v_{esc}^2}{v_0^2}}\right)
\end{equation}
\begin{figure} [h]
    \centering
    \includegraphics[width=12 cm]{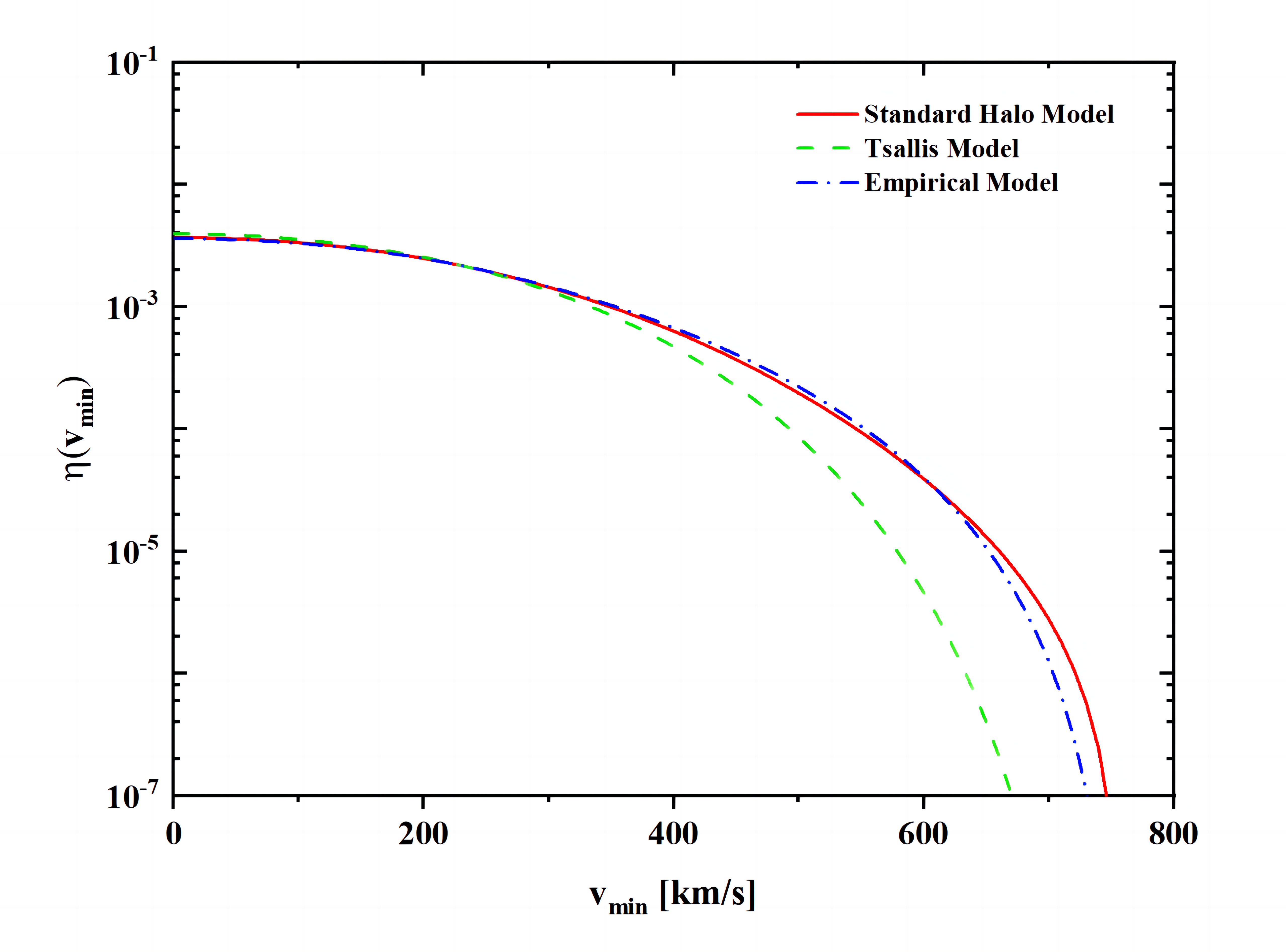}
    \caption{The $\eta(v_{min})$ for the three above-mentioned models as a function of the minimum velocity $v_{\rm min}$. The green dashed line represents the $\eta(v_{min})$ for the Tsallis Model, the red curve shows that for the Standard Halo Model, and the blue dot-dashed line illustrates that for the Empirical Model. 
    }
    \label{fig:eta_vmin}
\end{figure}
where $v_{esc}=528$ km/s is the escape velocity~\cite{Smith:2006ym,Piffl:2013mla,Monari:2018ckf,Deason2019TheLH,Radick:2020qip}, $v_{E}=232$ km/s is the Earth's Galactic velocity~\cite{Lee:2013xxa,McCabe:2013kea}, and $v_0=228.6$ km/s is typical velocity of Maxwell-Boltzmann distribution~\cite{Kuhlen:2009vh}. $K$ is the normalization coefficient that makes the velocity distribution function satisfy $\int f_{SHM}(v)\mathrm{d}^3v=1$ and the $\Theta$ function is a step-function. Secondly, the Tsallis Model is proposed to explain the data of the N-baryon numerical simulation and non-extensive systems. The DM velocity distribution in the Tsallis Model is given by
\begin{equation}
    f_{Tsa}(\vec v)\propto \left\{
\begin{array}{cc}
[1-(1-s)\frac{\vec v ^2}{v_0^2}]^{1/(1-s)}   & |\vec v|<v_{esc} \\
0         & |\vec v|\geq v_{esc}
    \end{array}\right.
    ,
\end{equation}
where $s=0.813$ is the entropic index~\cite{Radick:2020qip}. Finally, the Empirical Model with a hydro-dynamical approach utilizes numerical techniques to study the behavior of DM components in the presence of baryons and has the following DM velocity distribution,
\begin{equation}
    f_{Emp}(\vec v)\propto \left\{
\begin{array}{cc}
e^{-|\vec v|/v_0 (v_{esc}^2-|\vec v|^2)^p}   & |\vec v|<v_{esc} \\
0         & |\vec v|\geq v_{esc}
    \end{array}\right.
    ,
\end{equation}
We take the index $p=1.5$ in the following calculation and consider the impact of Earth's Galactic velocity in the latter two models. 


\begin{figure} [h]
\centering
\includegraphics[width=16cm]{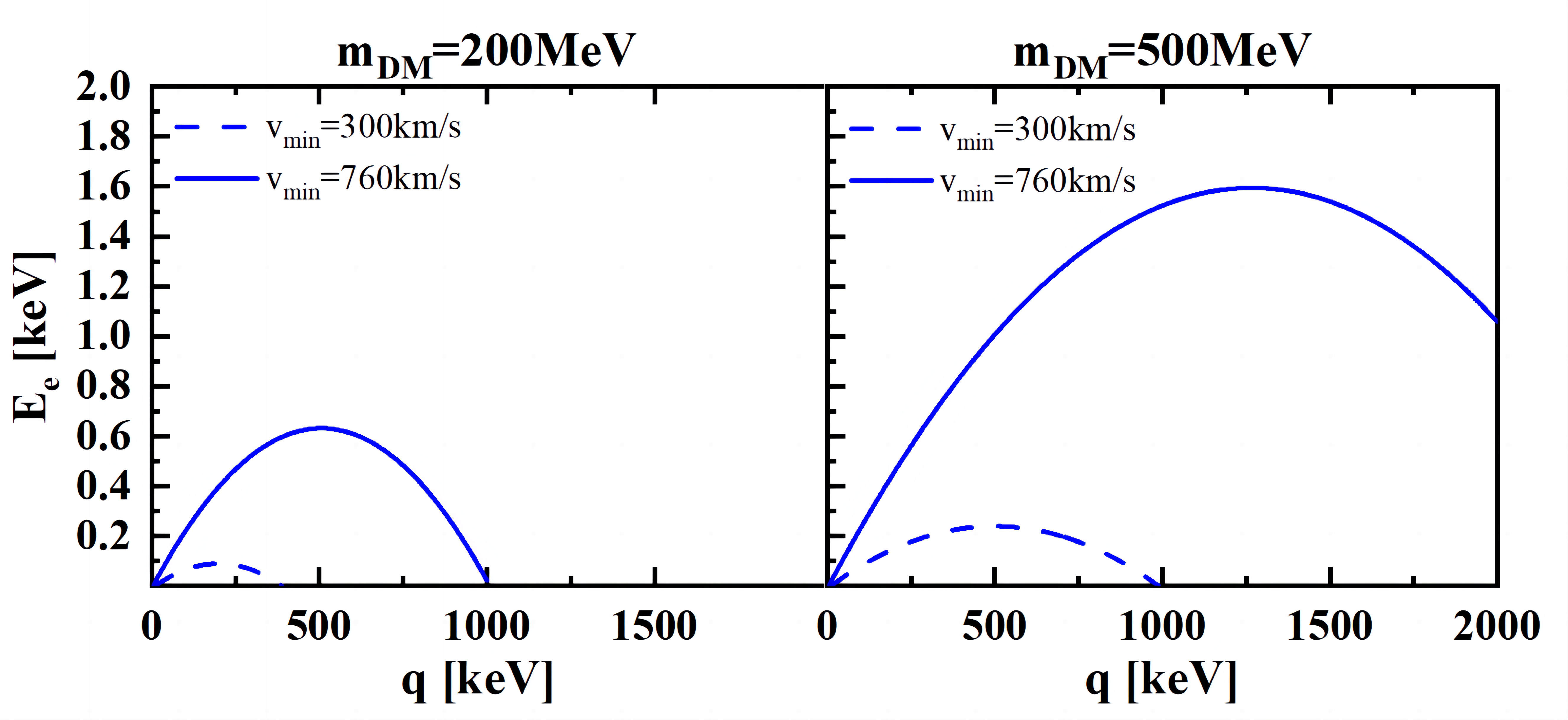}
\caption{
Contour plot of $v_{min}$ as a function of transfer momentum $q$ and the electron recoil energy $E_e$. In each panel, the dashed lines represent $v_{min}$=300 km/s, while the solid cruves illustrate $v_{min}$=760 km/s.}
\label{fig:vmin contour}
\end{figure}
The $\eta(v_{min})$ of the three different DM velocity distributions as a function of $v_{min}$ are shown in Fig.~\ref{fig:eta_vmin}. We can see that the Empirical Model and the Standard Halo Model start to have a tiny difference at $v_{min} \approx$ 650 km/s, while the Tsallis Model begins to diverge significantly from two other models at $v \approx $ 300 km/s.  To gain insight into the impact of the different DM velocity distributions on Eq.~\ref{eq2}, we draw a set of contour lines of $v_{min}$ derived by Eq.~\ref{vmin} in Fig.~\ref{fig:vmin contour}.

Fig.~\ref{fig:vmin contour} shows the contour plots of $v_{min}$ with the binding energy $|E_{n,l}|=10$ eV and two different DM masses $m_{DM}=200,500$ MeV. Note that the integral region of $q$ and $E_{e}$ is the area where the velocity of the incoming DM $v_{min} \leq v \leq v_{max}=v_{esc}+v_{E}=760$ km/s. As shown in Fig.~\ref{fig:vmin contour}, the integral from the $v_{min}=300$ km/s to $v_{max}=760$ km/s occupies the most $q$ and $E_{e}$ region. Evidently, the $v_{min}$ as a function of the transfer momentum $q$ and electron recoil energy $E_e$ has its lower bound $\sqrt{2(E_e + |E_{n,l}|)/\mu_N}$ when  $q = \sqrt{2\mu_N (E_e + |E_{n,l}|)}$. Additionally, the DM-nucleus reduced mass is approximately equal to $\mu_N \approx m_{DM}$ due to $m_{DM} \ll m_N$. Consequently, we can derive the lower bound of $v_{min} = \sqrt{2|E_{n,l}|/m_{DM}}$ by considering electron recoil energy $E_e=0$. With $|E_{n,l}|=10$ eV and $m_{DM}<20$ MeV, the minimum value of $v_{min}$ will always be greater than 300 km/s, at which the Standard Halo Model and Empirical Model are quite different from Tsallis Model. Therefore, the dependence of different DM velocity distributions on the differential event rates is worth being considered.

\begin{figure} [h]
    \centering
    \includegraphics[width=16 cm]{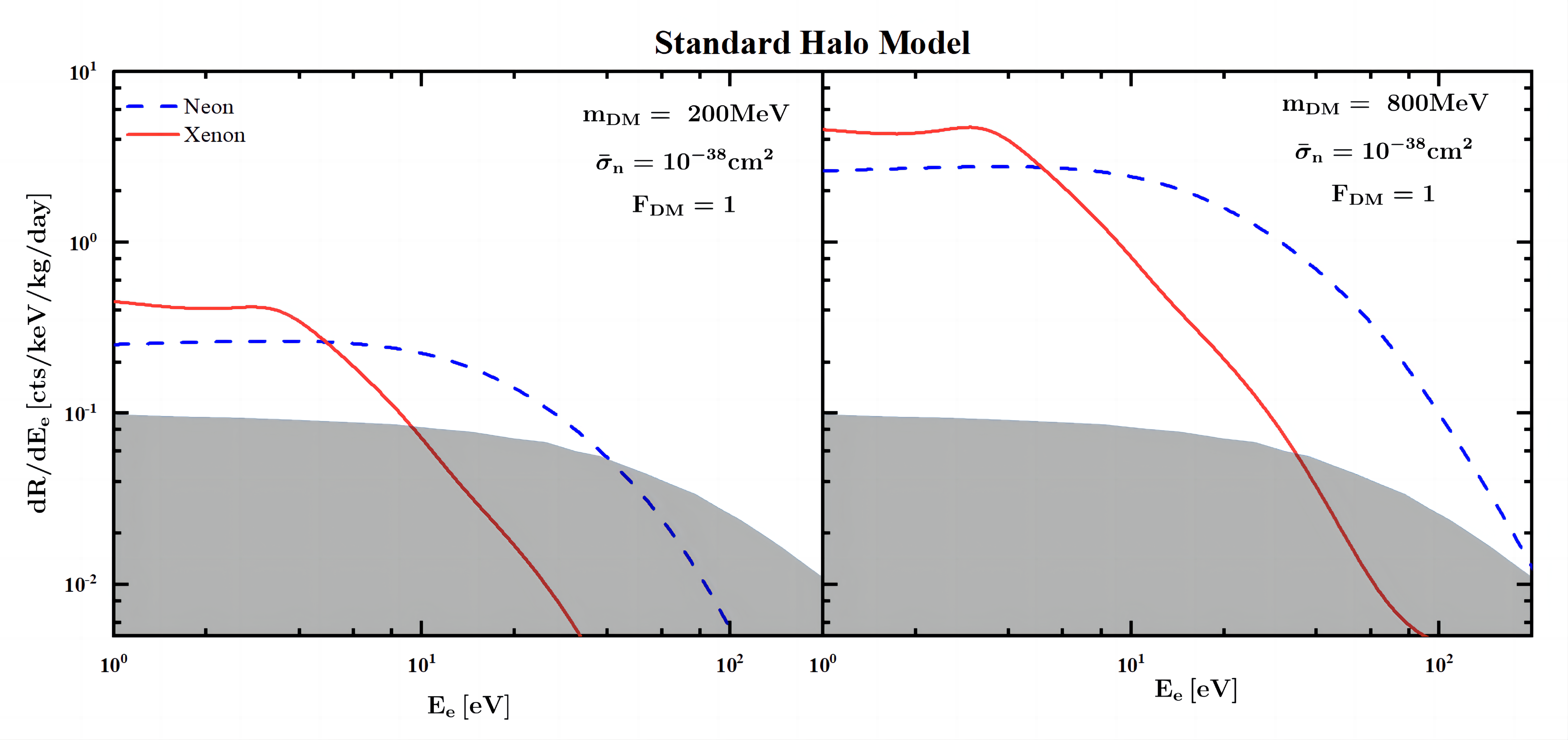}
    \caption{
    The differential events induced by Migdal effect with two different target materials (Xenon and Neon) versus the electron recoil energy $E_{e}$ for two different DM masses $m_{DM}=200$ MeV, 800 MeV, the reference cross section $\bar{\sigma}_n=10^{-38}$ cm$^2$ and DM form factor $F_{DM}=1$. In both left and right panels, the red solid and blue dashed lines are the results of Xenon and Neon as target materials respectively. The gray shaded region is the estimated background for DARKSPHER~\cite{Hamaide:2021hlp} provided that it is located at the Large Experimental Cavern at the Boulby Underground Laboratory.
    }
    \label{fig:dRdEe SHM}
\end{figure}
\section{NUMERICAL RESULTS AND DISCUSSIONS}
\label{sec4}

With the ionization form factors and the DM velocity distributions in hand, we can utilize Eq.~\ref{eq1} to calculate the differential events induced by DM-nucleus Migdal effect.
\begin{figure} [h]
    \centering
    \includegraphics[width=16 cm]{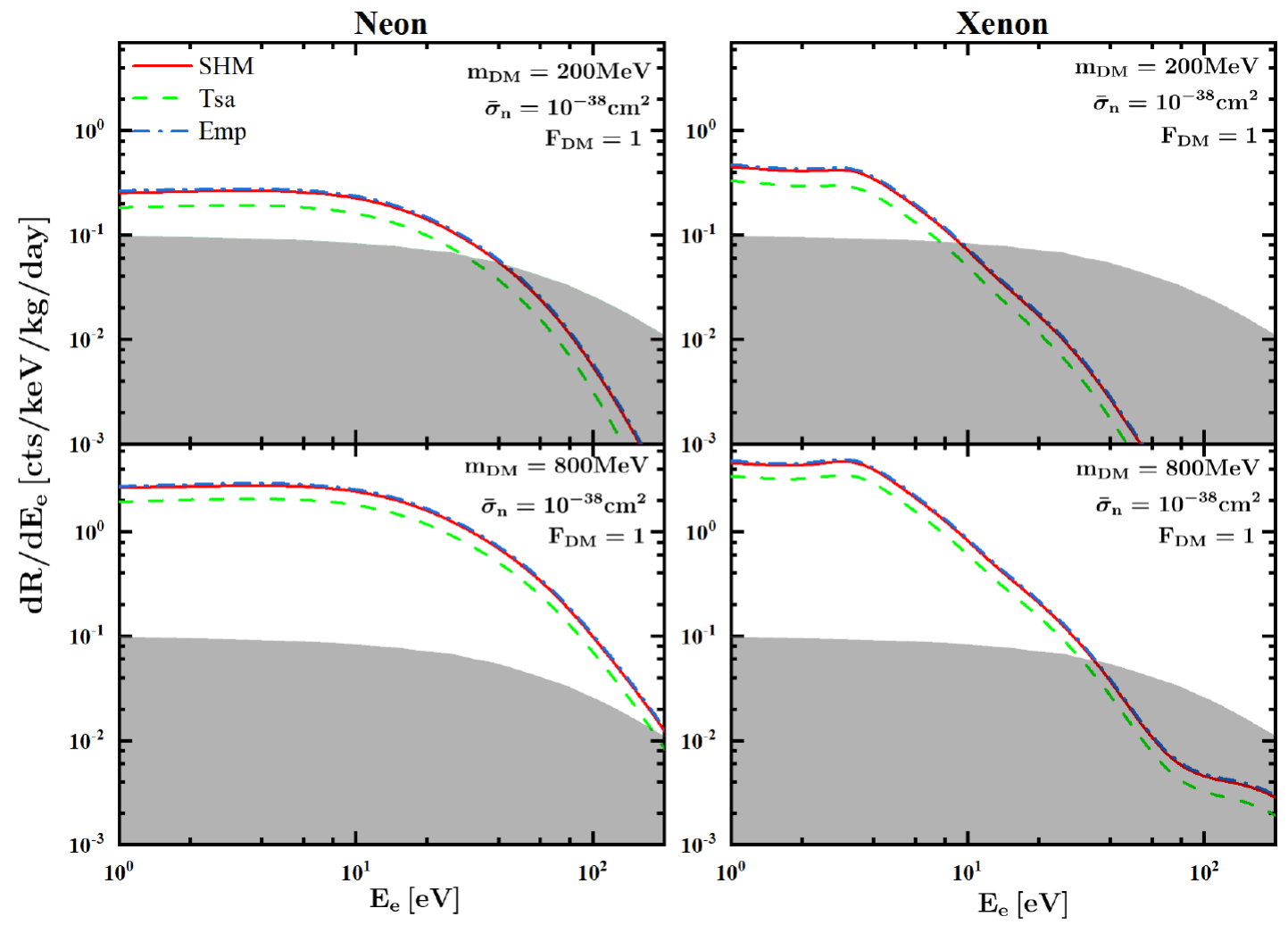}
    \caption{
    The differential events caused by Migdal effect for two different targets (Xenon and Neon) and three different DM velocity distributions as mentioned before. The other parameters are fixed the same as shown in Fig.~\ref{fig:dRdEe SHM}. The red, green dashed, and blue dot-dashed lines are the three DM velocity distributions of the Standard Halo Model, Tsallis Model and Empirical Model respectively.
    }
    \label{fig:comparison of dRdEe}
\end{figure}
The differential events for Xenon and Neon targets are shown in Fig.~\ref{fig:dRdEe SHM}, where we assume that the DM velocity distribution satisfies SHM. For Neon (Xenon) target, we consider that the electrons occupying the 1s, 2s, 2p(4d, 5s, 5p) energy level are ionized. As shown in Fig.~\ref{fig:dRdEe SHM}, the differential events for both Xenon and Neon targets decrease with the electron recoil energy $E_{e}$ increasing since the ionization factor $|f^{ion}_{n,l}(E_e,q_e)|$ is suppressed by large $E_{e}$. Furthermore, although they are greatly enhanced by $1/m_{T}$, the differential event rates induced by Xenon target are larger than those generated by Neon target in the small $E_{e}$ region. However, that is opposite in the large $E_{e}$ region. Due to the fact that the binding energy (12.7 eV) of the outermost level 5p of Xenon is lower than that of Neon (21.7 eV), the ionization factor $|f^{ion}_{n,l}(E_e,q_e)|$ for Xenon 5p energy level is much larger than that for Neon 2p energy in small $E_{e}$ region. While the total ionization factor for Neon is dominated over that for Xenon in the large $E_{e}$ region. This implies that Xenon as the target material has better performance for lower experimental thresholds. Besides, compared with the light DM, the heavy DM generates more events for the same target, DM form factor $F_{DM}$ and cross section $\bar{\sigma}_{n}$. This is because the heavy DM has more kinetic energy to induce more electron ionization. It should be noted that there exists a small peak around $E_e \sim 2.5$ eV in the differential events of the Xenon target, which arises from the small peak in the ionization function $|f^{ion}_{n,l}(E_e,q_e)|$ of the Xenon 4d energy level.


Fig.~\ref{fig:comparison of dRdEe} shows the impact of three different DM velocity distributions on the DM-induced differential events. The two upper panels represent $m_{DM}=200$ MeV, whereas the two bottom panels show $m_{DM}=800$ MeV. As shown in Fig.~\ref{fig:comparison of dRdEe}, the DM-induced differential events of the Standard Halo Model and the Empirical Model are almost the same because the DM velocity distribution of these two models has no significant difference in our interested region. On the contrary, the differential events generated by Migdal effect in the Tsallis Model are pretty different from those in the other two models. This is because that the velocity distribution of Tsallis Model are quite different from the Standard Halo Model and the Empirical Model in the large velocity region as shown in Fig~\ref{fig:eta_vmin}. Therefore, we mainly take into account the difference between the Tsallis Model and the Standard Halo Model.

We will present the exclusion limits on the $m_{DM}-\bar{\sigma}_n$ panel for NEWS-G projections. For electronic interactions, the quenching factor $Q$ is equal to 1~\cite{Bell:2019egg, Hamaide:2021hlp}. The ionization quenching effect in the Migdal effect is caused by the nuclear recoil. Since the quenching factor~\cite{Schumann:2019eaa} is dependent on the detector parameters such as the impurities, density of the medium and the electric field, it needs to be measured at a real detector condition. To estimate the quenching effect, one can use SRIM package~\cite{Katsioulas:2021pux,NEWS-G:2021mhf} to simulate the transport of ions in matter. For the Xenon target, we can take a constant quenching factor $Q=0.15$ for Migdal effect as Ref.~\cite{Bell:2021zkr}. While for the Neon material, we can parameterize it as follows~\cite{NEWS-G:2021mhf},
\begin{equation}
    Q(E_{\rm{nr}})=\alpha E_{\rm{nr}}^{\beta}
\end{equation}
where the parameters are $\alpha=0.2801$ and $\beta=0.0867$. We apply this ionization quenching factor $Q(E_{\rm{nr}})$ to calculate the differential events for different energy levels. By calculation, we find that for both Xenon and Neon gases, the differential events induced by the ionization quenching factor $Q=0.15$ and $Q(E_{\rm{nr}})$ are almost the same as those generated by $Q=0$. This implies that the nuclear ionization quench effects are negligible in our interested DM mass range. Additionally, we derive conservative constraints on $\bar{\sigma}_n$ via ignoring the ionization quenching effect induced by nuclear recoil energy $E_R$.
The primarily ionized electrons in the SPC detector will drift toward the anode in the cavity under the high-voltage electric field. During this process, we should consider the primarily produced electrons to create extra electron-ion pairs, which is related to the W-value of the filled gas. The W-value is the mean energy to create an extra electron-ion pair in media, for Xenon $W \sim$ 22 eV~\cite{10.2307/3575293}, and for Neon $W \sim$ 37 eV~\cite{doi:10.1063/1.1678247}. Ideally, the electron recoil $E_e$ should be converted to the experimentally observable electron-ion pairs. However, there is no exact description of the detector response at very low energy for NEWS-G experiment~\cite{Hamaide:2021hlp}. Although the conversion to the observables is of great importance for DM direct detection, this is not available for Xenon and Neon. Therefore, we take the same method as phenomenology in Ref.~\cite{Hamaide:2021hlp} to conservatively calculate the number of observable events induced by Migdal effect. According to the W-value of Xenon and Neon, we set the experimental threshold $E_{th}=$30 eV, which mimics a two-electron threshold. In addition, we also set experimental threshold $E_e=1$ eV to mimic a single-electron search threshold, which implies that the primarily ionized electron recoil energy is too small to produce additional electron-ion pairs in media. Since the DARKSPHERE detector proposed by the NEWS-G collaboration is scheduled to be put into operation in 2025, there is no available data from NEWS-G experiment. Instead, we simply integrate over the electron recoil energy $E_{e}$ to obtain the events induced by Migdal effect. Note that the DARKSPHERE detector with a 3-meter diameter and 5 bar pressure of the filled gas runs for a total of 300 days so that the total exposures of Neon and Xenon in DARKSPHERE detector are 48.08 ${\rm kg}\cdot {\rm year}$, 312.986 ${\rm kg}\cdot {\rm year}$ respectively. Additionally, the background events are indicated by the shaded region as shown in Fig.~\ref{fig:dRdEe SHM}. We can achieve the 90\% C.L. bound on the $m_{DM}-\bar{\sigma}_n$ plane by analyzing the signal and background events~\cite{McKeen:2018pbb},
\begin{equation}
\label{90}
    \frac{\Gamma(B+1,S+B)}{B!}=0.1
\end{equation}
where $\Gamma$ is the incomplete gamma function. $S$ is the total number of signal events derived by integrating over the electron recoil energy $E_{e}$ and $B$ is the total number of the background events. 

\begin{figure} [h]
\centering
\includegraphics[width=16cm]{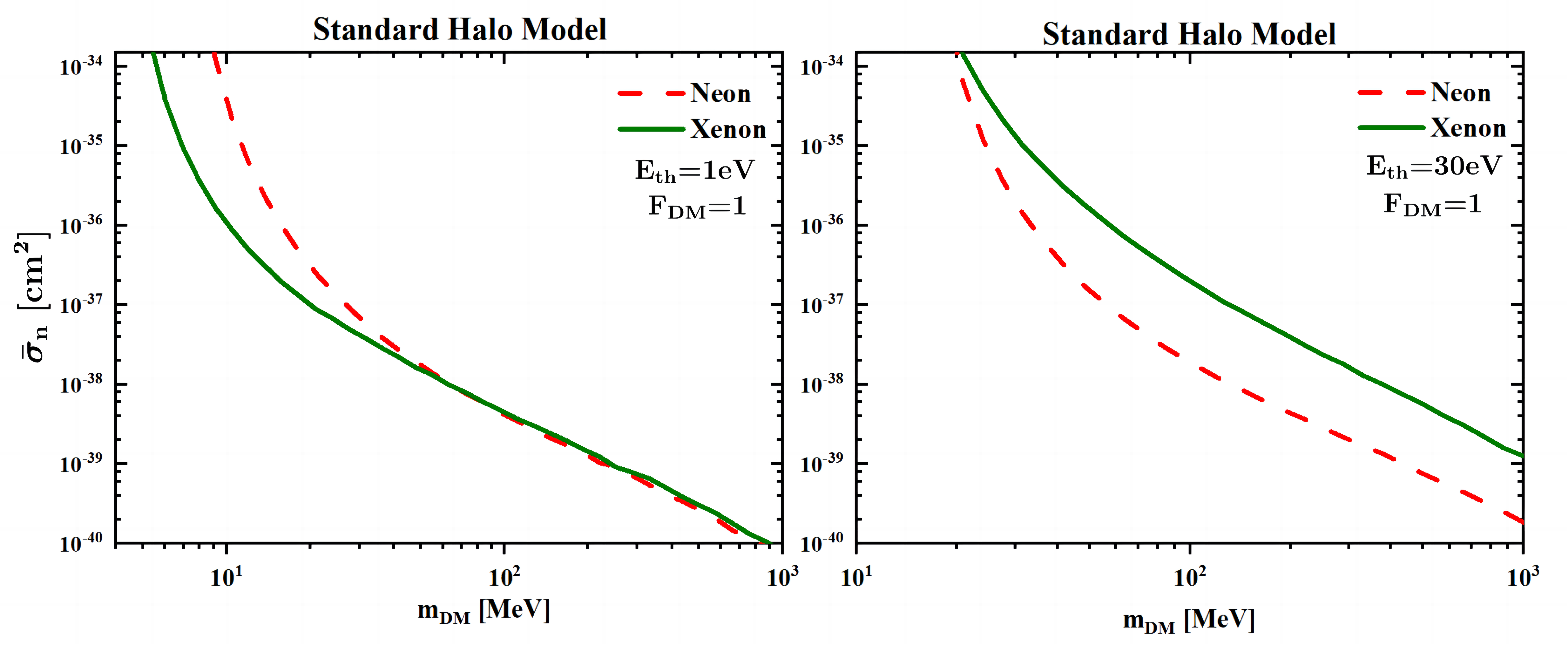}
\caption{
    Projected 90\% C.L. exclusion limits for the DARKSPHERE detector with two experimental threshold $E_{th}=1$ eV(left panel) and $E_{th}=30$ eV(right panel) on the $m_{DM}-\bar{\sigma}_{n}$ plane. In both left and right panels, the DM form factor $F_{DM}=1$ and the velocity distribution of DM is the Standard Halo Model. The red dashed lines represent the Neon target material while the green solid curves show the Xenon target material.
    }
    \label{fig:Eth}
\end{figure}

Fig.~\ref{fig:Eth} gives the 90\% C.L. exclusion limits for two different target materials on the $m_{DM}-\bar{\sigma}_{n}$ plane. Compared with the Neon target, the Xenon target gives stronger constraints on the $m_{DM}<50$ MeV region when the experimental threshold $E_{th}=1$ eV. As mentioned before, the binding energy of the energy level 5p for Xenon is smaller than that of the energy level 2p for Neon. Apart from overcoming the binding energy, the light DM has more kinetic energy to induce more observed events for Xenon target. Therefore, the reference cross section $\bar{\sigma}_{n}$ is strongly constrained by the low-threshold Xenon target, which can be used to probe the lighter DM. While the Neon target puts the more stringent limits on $\bar{\sigma}_{n}$ for the experimental threshold $E_{th}=30$ eV. This is mainly because the total ionization factor of Neon is much larger than that of Xenon in large electron recoil energy $E_{e}$ region. Besides, different from the Neon target, the differential events for Xenon target will be strongly suppressed by $1/m_{T}$. Therefore, with regard to the large experimental threshold ($E_{th}=30$ eV), many events will be induced by the Migdal effect for Neon target, resulting in the rigorous constraints on $\bar{\sigma}_{n}$. In addition, both Neon and Xenon targets simultaneously lose sensitivity at $m_{DM} \sim 20$ MeV in the right panel since the light DM with small kinetic energy cannot overcome the binding energy of the outermost electrons for both Neon and Xenon. Combined with the previous discussion, Fig.~\ref{fig:Eth} implies that the DARKSPHERE detector should be filled with the best performance gas for various thresholds in order to optimize experimental performance. 

\begin{figure}[ht]
    \centering
    \includegraphics[width=16cm]{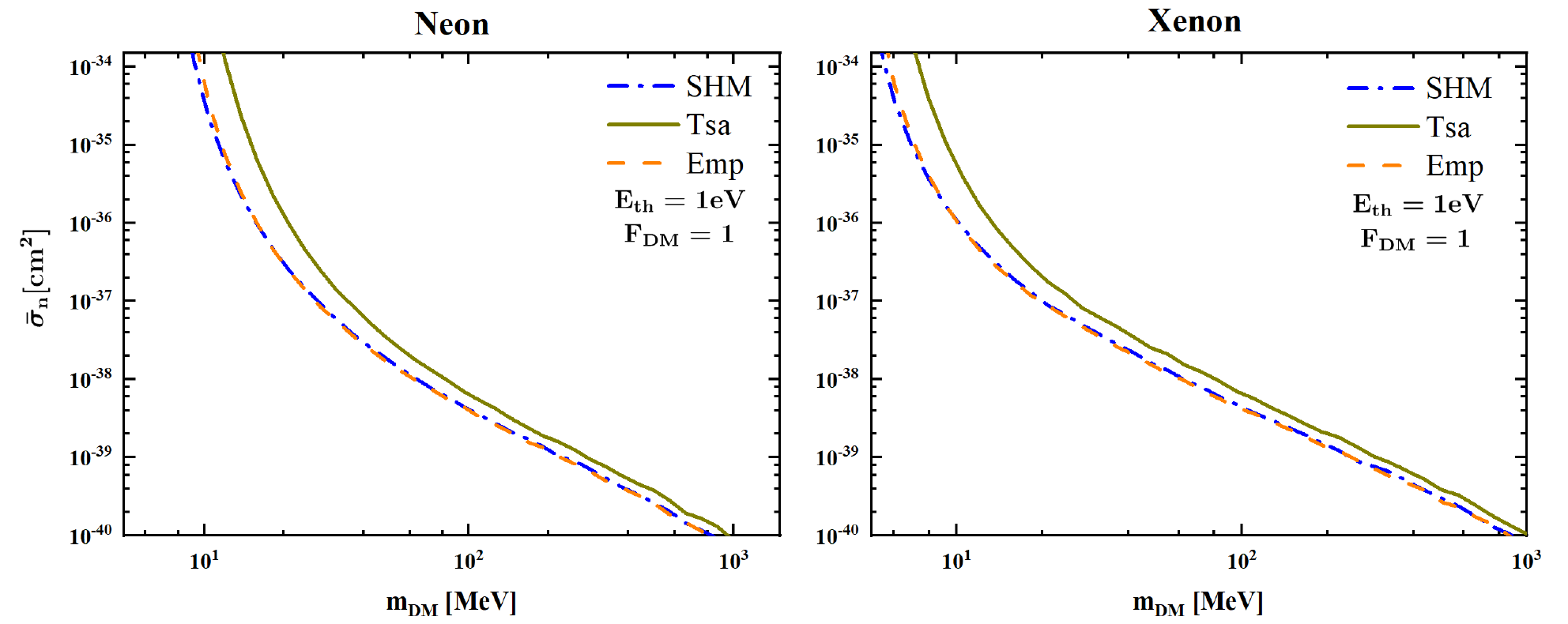}
    \caption{ The 90\% C.L. exclusion limits on reference cross section $\bar{\sigma}_{n}$ versus $m_{DM}$ for three different DM velocity distributions $f({v_{DM}})$, DM form factor $F_{DM}=1$ and the experimental threshold $E_{th}=1$ eV. The three different DM velocity distributions includes Standard Halo Model (the blue dot-dashed line), Tsallis Model (the dark yellow solid line), and Empirical Model (the orange dashed line). The target materials are the Neon (left panel) and Xenon (right panel).
    }
    \label{fig:comparison of sigma}
\end{figure}
Additionally, the dependence of exclusion limits on the three DM velocity distributions is shown in Fig.~\ref{fig:comparison of sigma}. As shown in Fig.~\ref{fig:comparison of sigma}, the exclusion limits derived by the Standard Halo Model and Empirical Model are almost the same, which are quite different from those caused by Tsallis Model, especially for the light DM. The light DM implies the large allowed minimum velocity $v_{min}$ where the velocity distribution of Tsallis Model is quite different from the other two models. It should be noted that for Xenon target, the exclusion limits on $\bar{\sigma}_{n}$ induced by light DM-nucleus scattering in the Standard Halo Model is an order of magnitude stronger than that originating from the Tsallis Model. This indicates that the dependence of various velocity distributions on DM direct detection should be considered, especially for light DM.

Given that all above calculations are based on $F_{DM}=1$, we should obtain the exclusion limits on $\bar{\sigma}_{n}$ by taking into account the other DM form factors~\cite{Bloch:2020uzh}:
\begin{itemize}
    \item [1)] $F_{DM}=1$, “heavy” mediator;
    \item [2)] $F_{DM}=(\frac{\alpha m_e}{q})^2$, "light" mediator;
    \item[3)] $F_{DM}=\frac{q}{\alpha m_e}$, $q$-dependent “heavy” mediator;
    \item [4)] $F_{DM}=(\frac{q}{\alpha m_e})^2$, $q^2$-dependent “heavy” mediator.
\end{itemize}
\begin{figure} [h]
\centering
\includegraphics[width=16cm]{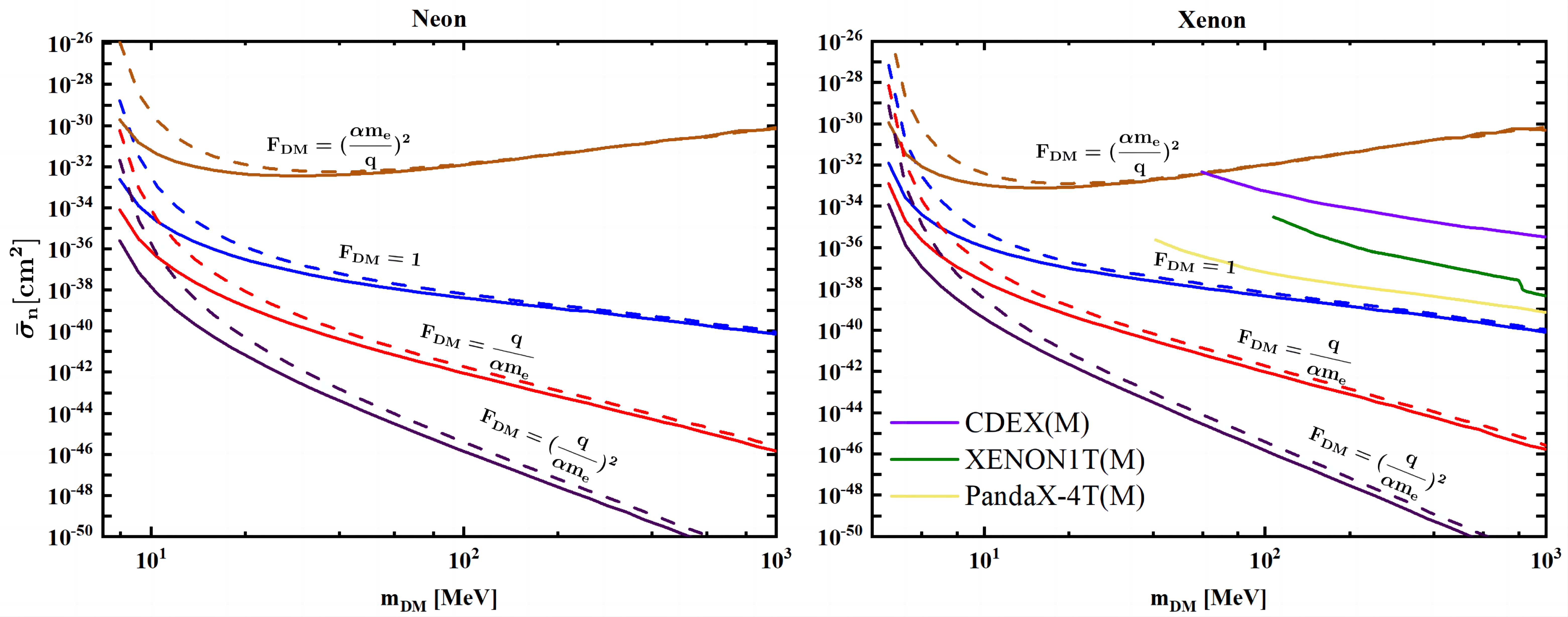}
\caption
    {The 90\% C.L. exclusion limits on the $m_{DM}-\bar{\sigma}_{n}$ for four different DM form factors $F_{DM}$ and the experimental threshold $E_{th}=1$ eV. The solid and dashed lines indicate the results generated by the Standard Halo Model and Tsallis Model respectively. The left (right) panel represents the Neon (Xenon) target material. The purple line shows the constraint from the CDEX experiment~\cite{CDEX:2019hzn} while the green line illustrates that from the XENON1T experiment~\cite{XENON:2019gfn,XENON:2019zpr}. The yellow line is the results of the PandaX-4T experiment by considering the Migdal effect with the mediator mass of 1 GeV~\cite{huang2023search}.}
    
\label{fig:Sigma compare}
\end{figure} 
The 90\% C.L. exclusion limits derived from Migdal effect by considering four different DM form factors $F_{DM}$ are delineated in Fig~\ref{fig:Sigma compare}. For both the Neon and Xenon targets, the reference cross section $\bar{\sigma}_{n}$ is the most weakly constrained when the DM form factor $F_{DM}=\left(\alpha m_{e}/q\right)^2$. While that is the most strongly limited when the DM form factor $F_{DM}=\left(q/\alpha m_{e}\right)^2$. As discussed before, the minimum of the transfer momentum $q_{-}$ is always larger than $\alpha m_{e} \sim 4$ keV. Because of being enhanced by the DM form factor $F_{DM} \propto q^2$, more events will be generated by the DM form factor $F_{DM}=\left(q/\alpha m_{e}\right)^2$ and receive stronger constraints. Whereas, fewer events will be produced by $F_{DM}=\left(\alpha m_{e}/q\right)^2$ due to being suppressed by $F_{DM} \propto 1/q^2$, which leads to weaker constraints. With regard to Xenon target, the exclusion limits arising from the other DM form factors except for $F_{DM}=1$ are stronger than those derived by the CDEX, XENON1T and PandaX-4T experiments. Also, compared with those from CDEX, XENON1T and PandaX-4T experiments, the exclusion limits derived by using the DARKSPHERE detector can reach the lighter DM mass region, which implies that the NEWS-G experiment has the potential to detect light DM. Additionally, the differences between two different DM velocity distributions are quite different in the light DM mass region as mentioned before, which results in orders of magnitude impact on reference cross section constraints. Especially, owing to the low experimental threshold of Xenon target, the constraints on cross section $\bar{\sigma}_{n}$ can reach $\bar{\sigma}_n \sim10^{-39}$ cm$^2$ for $F_{DM}=\left(q/\alpha m_{e}\right)^2$when $m_{DM} \sim 10$ MeV, which are stronger than those derived from existing XENON1T and CDEX experiments.
\section{CONCLUSIONS}
\label{sec5}
The spherical proportional counter proposed by the NEWS-G collaboration for DM direct detection has the property that it can be flexibly filled with various experimental target materials, such as Neon and Xenon. The NEWS-G experiment can probe the DM mass $m_{DM}$ being down to as low as sub-GeV. In this work, we derive the exclusion limits on the $m_{DM}-\bar{\sigma}_{n}$ by exploiting the DARKSPHERE detector after considering the Migdal effect with different target materials, DM velocity distributions $f\left(v_{DM}\right)$ and form factors $F_{DM}$. We find that the exclusion limits depend on the DM velocity distributions, especially for the light DM. For DM mass $m_{DM} \sim {\cal O}(10)$ MeV, the difference of exclusion limits on reference cross section $\bar{\sigma}_n$ between the Standard Halo Model and Tsallis Model can reach orders of magnitude. Because of the low binding energy of 5P energy level for Xenon, the DARKSPHERE detector with the low experimental threshold ($E_{th}=1$ eV) can probe the light DM with mass being down to ${\cal O}(10)$ MeV. However, the exclusion limits derived by the DARKSPHERE detector with the relatively high experimental threshold ($E_{th}=30$ eV) for Neon target are stronger than those for Xenon target. Therefore, the DARKSPHERE detector should be filled with the best performance gas for various thresholds in order to optimize experimental performance. The spherical proportional counter with suitable filled gas has the potential to detect sub-GeV or even lighter DM through Migdal effect. Besides, with regard to different DM form factors $F_{DM}$, the constraints on the reference cross section $\bar{\sigma}_n$ for $F_{DM}=\left(q/\alpha m_{e}\right)^2$ are the strongest because of the caused events being enhanced by $q^2$ while those for $F_{DM}=\left(\alpha m_{e}/q\right)^2$ are weakest due to the generated events being suppressed by $q^2$. Given that the low experimental threshold for Xenon target, the constraints on cross section $\bar{\sigma}_{n}$ can reach $\bar{\sigma}_n \sim10^{-39}$ cm$^2$ for $F_{DM}=\left(q/\alpha m_{e}\right)^2$ when $m_{DM} \sim 10$ MeV, which are stronger than those derived from existing CDEX, XENON1T and PandaX-4T experiments.

\section*{Acknowledgements}
\label{sec6}
We thank Christopher McCabe for helpful discussions on the NEWS-G experiment, Masahiro Ibe, Wakutaka Nakano and Yutaro Shoji for helpful discussions on the usage of FAC, . We also acknowledge Liangliang Su for useful discussions. This work is supported by the National Natural Science Foundation of China (NNSFC) under grant No. 11805161 and 12147228, by Natural Science Foundation of Shandong Province under the grants ZR2018QA007. 

\bibliography{refs}

\end{document}